\documentclass[useAMS,usenatbib]{mn2e}
\usepackage{amssymb}
\usepackage{graphicx}

\title[Axis change in AGN black hole]{B0707-359: a case study of change in AGN-black hole spin axis}
\author[Saripalli et al.]
  {L.~Saripalli,$^1$\thanks{E-mail: lsaripal@rri.res.in}
  J. M.~Malarecki,$^2$ \newauthor R.~Subrahmanyan,$^{1,3}$   D.~H.~Jones,$^4$ L.~Staveley-Smith,$^{2}$\\
   $^1$Raman Research Institute, C V Raman Avenue, Sadashivanagar, Bangalore 560080, India\\
   $^2$International Centre for Radio Astronomy Research, M468, The University of Western Australia, Crawley, WA 6009, Australia\\
  $^3$Science Operations Center, National Radio Astronomy Observatory, P O Box 0, Socorro, NM 87801, USA\\
  $^4$School of Physics, Faculty of Science, Monash University, Clayton, Victoria 3800, Australia\\
  }

\begin{document}

\date{Accepted 2013 August 22. Received 2013 August 22; in original form 2013 April 26}

\maketitle

\label{firstpage}

\begin{abstract}
Structures of radio galaxies have the potential to reveal inconstancy in the axis of the beams, which reflect the stability in the spin axis of the supermassive black hole at the center.  We present radio observations of the giant radio galaxy B0707$-$359 whose structure offers an interesting case study of an AGN that may be exhibiting not only inconstancy in AGN output but also inconstancy in direction of ejection axis. Its radio morphology shows evidence for a restarting of the jets accompanied by an axis change.  The observed side-to-side asymmetries of this giant radio galaxy suggest that the new jets are not in the plane of the sky. We infer that the hotspot advance velocities are unusually large and of magnitude a few-tenths of the speed of light.  The dual-frequency radio images are consistent with a model where the beams from the central engine ceased, creating a relic double radio source; this interruption was accompanied by triggering of a movement of the axis of the central engine at a rate of a few degrees~Myr$^{-1}$.   The closer location of the giant radio galaxy axis to the host minor axis rather than the host major axis is supportive of the restarting and axis-change model for the formation of the double-double structure rather than the backflow model.
\end{abstract}


\section{Introduction}
Giant radio galaxies and radio galaxies exhibiting evidence of restarted activity, the so-called 'double-double' radio galaxies are both testimony to the capability of jets and black hole spin axes to maintain long-term steadiness in their orientation. 
On the other hand, there has been evidence as well as several suggestions of precessing jets and of jets that might have 
flipped in orientation. Axis precession and flips may happen because of the influence of a nearby galaxy or as a consequence of a binary black hole merger \citep{eke78, mer02, cam07}. That jets can precess is already evident in the well-studied 
galactic micro-quasar SS433 and inferred in radio galaxies such as 3C123 (also 3C294, \citet{erl06}). However, though simulations of binary black hole mergers have revealed that flips in jet axes are plausible mechanisms, there are few conclusive observational examples where an axis flip may have occurred (e.g. the triple double source 3C293 \citep{chi96} and 4C+00.58 \citep{hod10}). Detecting a flip in jet axis via morphology requires that the source be active along a previous axis for long enough time to have created extended lobes that survive as distinct relics that are visible in radio or X-ray images. Although morphologies of X-shaped radio galaxies may suggest occurrence of such an axis change there are several problems with this inference as being a general formation mechanism for this class of sources \citep{cap02, sar09, hod11}. Examples of radio galaxies
with radio morphological evidence for changed axes between activity epochs are rare.  More discoveries of potential examples where axes have changed over time are needed since they form a valuable resource for understanding the occurrence rate of perturbations to the black hole axes and for examining timescales and conditions under which the axes change.

We present radio observations of the 1.10-Mpc giant radio galaxy B0707$-$359 (redshift, z=0.1109; \citet{mal13}. The radio source is an unusual giant 
in that it represents a case of an AGN that appears to have restarted with a changed axis. We have imaged this source in the radio
as part of a study of the ambient intergalactic medium using a sample of giant radio galaxies \citep{mal13}.  We adopt a flat cosmology with Hubble constant $H_{0}$ = 71~km~sec$^{-1}$~Mpc$^{-1}$ and matter 
density parameter $\Omega_{m}$ = 0.27.

\section{Observations and Results}

The radio source was observed simultaneously at frequencies 1.4 and 2.3~GHz in several array
configurations over several days using the Australia Telescope Compact Array (ATCA). 
A log of the observations is given in Table~1. The observations were carried out in total intensity and
polarization. The observations of the source were interspersed with short 1.5 to 4~min observations
of a suitable secondary calibrator B0727$-$365. The flux densities were set based on the known flux density
of the primary calibrator B1934$-$638, which was observed once at the start  of the observing run.
The source was observed for a total of 53.4~hr. Standard data
reduction procedure was followed using MIRIAD in obtaining the final images.  The data was
phase self-calibrated to improve the imaging dynamic range. The final images are 
presented in Figs.~1-3.

The total intensity maps at  1384 and 2368 MHz are shown in Figs.~1 and 2; the radio morphology is a remarkable double-double structure: twin, very dissimilar lobe pairs that are skewed
with respect to each other by as much as 30 degrees.  The outer lobe pair is asymmetric in extent from the center and the lobes are dominated by bright hotspots at their ends. The second inner lobe pair is shorter in extent, diffuse and much fainter in surface brightness and quite symmetric about the core.  These inner lobes are  edge-brightened; however, they show no evidence of hotspots in either lobe. 


The polarization distributions are shown in both figures overlaid as vectors representing the electric field.  The vectors have been corrected for Faraday rotation, which was estimated using the polarization data at 1384 and 2368 MHz. The polarization data reveals both pairs of lobes to be strongly polarized: the diffuse lobes have a mean percentage polarization between 10 and 20 per cent. In these lobe regions the projected magnetic field vectors are oriented parallel to the lobe edges. 

Fig.~3 shows the spectral index distribution over the source. The spectral index distribution was computed using the 
1384~MHz total intensity image and a separate low-resolution 2368~MHz image, which was produced from the calibrated data described  above with the spatial frequency range truncated to a maximum of 28~k$\lambda$ so that the resulting image is matched in resolution to that of the 1384~MHz image. The limiting of the visibility coverage resulted in a reduction of the angular resolution from the 12~arcsec of Fig.~2 to 16~arcsec. The reduced resolution image was separately self-calibrated to improve the image dynamic range. The 2-point spectral index  map shows a generally steepening spectral index distribution towards the core region in the NW inner lobe.
The hotspots at the ends of the outer lobes have the expected flatter spectra.  The core has an inverted spectrum ($S_{\nu} \propto \nu^{\alpha}$; $\alpha= 0.44$), which is not uncommon in radio galaxies \citep{sle94}.

In Table~2 are listed some observed and derived properties for the source.

\section{Discussion}

The relatively flat spectrum hotspots at the ends of the lobes are clearly the sites of termination shocks where jets from the center currently end.  The open question is the past evolution of the jets and their axes, which has implications for jet continuity and axis stability.  Did the source experience an abrupt increase in hotspot advance speed, leaving behind the diffuse inner lobes  and rapidly advancing to form the current hotspots? Do the inner lobes represent backflow of relativistic plasma from the sites of current hotspots, that are now along an axis offset from the current jet axis because the backflow has been deflected by the thermal halo associated with the host galaxy?  Or do the inner lobes represent relic relativistic plasma that was deposited at their current sites by jets that were along the axis of the inner lobes in a previous epoch of activity?  In the subsections below we consider these alternatives in the light of the observations presented here.

The giant radio galaxy has prominent hotspots at the ends and has lobes that remain relatively narrow over some distance towards the core before flaring in
opposite directions.  The NW and SE inner regions may be interpreted as flared parts of the radio lobes, which are recessed and confined to the vicinity of the core, with the outer regions viewed as protuberances.   In this picture, the lobes on both sides have a bottle-neck appearance. Such a morphology has been observed in several prominent  3C radio galaxies (e.g. 3C33, 3C79, 3C132, 3C184.1, 3C234, 3C349, 3C381 and 3C390.3). Such sources with protruding hotspots have been speculated to form because of either a sudden drop in the ambient medium density or sudden increase 
in the jet density or power or in the steadiness of the beam (\citet{lea91, har97}); these circumstances result in a step increase in advance speed of the hotspot.

As we trace the outer lobes from the hotspots towards the core, the flaring in the form of the inner lobes first occurs at similar distances of about 250~kpc  from the core. These flared regions are very similar in appearance and extend in opposite directions. They are collinear 
with the core forming an axis which is offset from the axis defined by the hotspots by an angle of about 30~degrees. 
They are both well-bounded at their 
leading ends and have projected magnetic field vectors that run parallel to the edges. All these characteristics  
of the flared regions suggest a resemblance in several respects to 
radio lobes rather than the central flares and distortions seen in radio galaxies \citep{lea84, sar09}. The two
emission regions also appear to lack compact features and they have a relaxed appearance with relatively low
axial ratios.

The giant radio galaxy B0707$-$359 displays a structure that has both inversion symmetry and a marked offset between the outer and inner lobes. Any explanation proffered for the radio structure we observe in B0707-359, whether in terms of a sudden drop in the ambient medium density or sudden increase in the jet density or power or in the steadiness of the beam, will need to take into account the offset in axes between the inner and outer lobes.

\subsection{Is the structure in B0707$-$359 caused by abrupt changes in ambient medium?}

For an abrupt change in density of the ambient medium to be responsible for the narrow protrusions of the two outer lobes, the density change will not only have to occur at the large distance of nearly 250~kpc from the host galaxy but the change will need to be such that it also deflects the two 
jets inversion-symmetrically in opposite directions. While density changes in the associated thermal halos of elliptical galaxies may occur symmetrically on both sides, they are unlikely to occur  at such large distance from the host. The change in ambient density is also required to occur at more than twice the distance over which bottle-neck structures are observed to appear in the other examples known in the literature. More significantly, the lack of abrupt bends in the two outer lobes (suggestive of deflection as they emerge out of the inner lobes) and the common axis they appear to share pose a difficulty for this model. It is more likely that changes in jet properties instead may be responsible for the peculiar
structure of B0707$-$359: either an increase in jet density (hence power) or in its steadiness, accompanied also by a change in the axis of the jets.  

\subsection{Is the structure in B0707$-$359 caused by deflection of backflows?}

Deflection of backflows in a thermal medium associated with the host galaxy has been suggested
as a mechanism by which inversion-symmetric central distortions in radio galaxies (type-2 distortion; Fig.~6(a);  \citet{lea84})
as well as the 'wings' in X-shaped radio galaxies (XRGs) may be produced (e.g. \citet{lea84,wor95}). Could the same physical process 
also be responsible for the offset inner lobes in B0707$-$359? Given the presence of prominent hotspots at the ends of the outer lobes 
strong backflows are expected. However, the backflows appear to stop well short of the core and instead deviate from the main axis 
at scales of more than two hundred kiloparsec from the core on either side. This is in sharp contrast to
the features commonly seen as central lobe distortions in samples of radio galaxies or the wings in XRGs. Moreover, the two inner lobes in 0707$-$359 
as already described are edge-brightened, well-bounded emission regions offset at a relatively small angle with respect to the main radio axis all of 
which also contrast with properties associated with commonly seen distortions in radio galaxies or XRGs. Given the strong differences the thermal halo may not
have a major influence on the backflows.

In the context of the reported geometrical relationship between the host optical axes and radio axes of radio galaxies \citep{cap02, sar09} 
where radio galaxies with both prominent central distortions and XRGs were found to predominantly lie closer to the host major axes it would be 
revealing to examine the relative orientations in B0707$-$359. Given the arguments presented against backflow deflection mechanism in the
formation of the offset structures in B0707$-$359 such a major-axis orientation of the radio axis is not expected. Instead, given the propensity
of giant radio galaxies to lie with their radio axes closer to the host minor axes \citep{sar09} it suggests a minor-axis orientation for B0707$-$359.

Using available data we have estimated the major axis position angle of the host galaxy. GALFIT (version 3.0.5) was used to fit the image of the B0707-359 optical host detected in DSS (red), 2MASS and Swift (V-band) data. 
Four Swift V-band datasets were summed to form a single image, with a combined exposure time of 1117s, using the HEASOFT programs UVOTIMSUM and XIMAGE. For each image, the plate scale in arc-seconds per pixel and photometric zero-point were set, and the sky background in the fitting region in ADUs and integrated magnitude for the host were estimated. The position angle of the host elliptical galaxy (as measured by GALFIT in degrees from north through east) is 78, 82 and 72 using the DSS, 2MASS and Swift data, respectively. While in all the datasets the host galaxy counts may be affected by light from a nearby bright star to the east, the 2MASS image has a relatively poor signal to noise ratio and may be less reliable. The derived major axis position angle (offset by nearly $55\degr$ with respect to the outer double radio axis, considering DSS and SWIFT data) agrees (1) with our reasoning above against formation of the offset lobes as a result of backflow deflection (where the radio axis is expected to be located closer to major axis) and (2) with the expectation of a closeness with host minor axis for the axes of giant radio galaxies  \citep{sar09}.

\subsection{Has there been an axis flip in B0707$-$359?}

The more simplistic scenario for the source structure we observe in B0707$-$359 is that the source may have experienced a
flip in the jet axis between two epochs of activity. The source was initially
active along the direction of the inner lobes and the jet activity then switched off for a period of
time after which it again became active but along a different direction that is offset by about 30 degrees to the
north west.  Restarting---the switching off and switching back on of the AGN beams---in radio galaxies is today a well accepted model that has been established to occur in several cases and has 
accumulated much support over the years \citep{bri86, roe94, chi96, sub96, sch00, ode01, sar02, sar03, sar05, mar06, 
sai06, jam07, bro07, jam09, mar09, blu11, sar12, gia12}.  Additional evidence comes from the examples of and formation scenarios for multiple pairs of X-ray `holes' associated with AGN at the centers of galaxy clusters and groups \citep{mcn07}. 

Our two-frequency spectral-index distribution unfortunately does not allow a clear picture to emerge. If the two inner
lobes are the relics of past activity, not only should they have steeper spectral indices compared to the rest of the 
source, they should also have spectral gradients with steeper spectral indices towards the central regions. Both inner lobes 
do have steeper spectral indices on average than the outer lobes, although only -0.9 compared to -0.8 in the outer lobes.
However, the spectral index distribution (Fig. 3) does not clearly show the expected steepening towards the
central regions; while NW inner lobe may have such a trend, such a behavior is not apparent in the case of the SE inner lobe. Regions of flatter
indices are noticed in both the inner lobes along the main radio axis (formed by the outer hotspots) possibly indicating influence of younger jet 
material associated with the outer lobes. Whereas further away from the common emission regions of both NW and SE inner lobes prominent areas of steep spectral indices of -1 to -1.2 are indeed seen.

The source morphology, polarization structure and to a lesser extent the spectral index distribution are all consistent with the
likelihood of a restarting of activity in this source where the new jets have emerged along a changed direction. 
We also note here that the integrated radio power of the inner lobe pair alone (Table~1) is comparable to that of radio galaxies, being close to the dividing line between edge-brightened and edge-darkened radio galaxies. 
Here we pause and ask
whether the inner double, with its low lobe axial ratio, lack of hotspots in an
edge-brightened structure and circumferential magnetic fields, may instead be a lobed FR-I like 3C~296 or NGC~326 rather than being the relic of a previous FR-II episode (and hence suggesting instead an axis change from outer to inner double). This possibility however is less likely because of absence of prominent twin-jets found ubiquitously in FR-I sources. 
Taken together with the short lifetimes of hotspots seen in the outer double (few $10^{4}$ yr) and the age of the weaker episode (powered by slower jets), 
the inner double is unlikely to be a lobed FR-I. This is supportive of the jets initially being located along the position angle of the inner double and then rotating to the axis of the outer double.

A change in the direction of the jets when restarting in a new epoch is not common. Most examples of radio galaxies with restarted 
activity show jet axes that have remained unchanged. The few examples where a changed jet direction is
recognized are the well known triple-double source 3C293 \citep{chi96} and the more recent example of 
4C+00.58 \citep{hod10}. In these examples the emission regions corresponding to the two epochs are distinct whether in
morphology or physical separation or stage of activity (e.g. in 3C293) lending a strong case for change of axis after 
a switching off of the central beams.

The morphology of the older epoch emission will depend on the time for which the
jets have remained switched off (the quiescent time).  In B0707$-$359,
if indeed the inner pair of lobes represent older activity, the quiescent time may not have been long because the lobes are well-bounded
over their 250~kpc extent although they lack hotspots and have low axial ratios. Using an equipartition magnetic field of about 2.5 micro-Gauss
\citep{mal13} for the inner lobes and the fact that the break frequency is above 2.3~GHz the upper limit to the source age is derived to be $10^{8}$ yr.

A pronounced asymmetry (of 1.47) is seen for the outer lobes which is not seen for 
the inner pair. Because of the rather similar appearance and symmetric extent of the inner lobe pair, we may expect that there is little asymmetry in the properties of the external medium over a distance of at least 250~kpc from the host galaxy. For the environment to be responsible for the asymmetry in the extents of the outer lobes, the medium on scales larger than 250 kpc must be asymmetric. While this is possible, we discuss below an alternate and more likely origin for the asymmetric appearance. The available optical images are not of sufficient quality to assess the neighbouring galaxy environment. This is being studied as part of a separate program involving a larger sample of giant radio galaxies (Malarecki et al. in preparation).

The alternate origin for the asymmetric appearance of the outer lobes involves light travel-time effects. It is 
likely that the new and current jet axis is not in the plane of the sky. A current jet axis with the approaching 
side towards the SE will result in a longer extent for that lobe compared to the opposite NW side because of 
light travel-time effects. For inclined jets to give rise to a pronounced asymmetry the 
advance speeds of the hotspots will need to be relativistic.
The optical spectrum of the host galaxy does not show broad emission lines \citep{mal13}, which suggests that the axis of the AGN is inclined by more than $45\degr$ to the line of sight.  To produce the observed asymmetry of 1.47 in the ratio of the apparent projected lengths of the two outer lobes, a head advance speed of at least 0.27$c$ is necessitated by the constraint on the inclination angle: larger and unrealistic advance speeds are required to explain the asymmetry ratio if the inclination angle is much greater than $45\degr$.  This argues for an inclination angle close to $45\degr$ along with advance speed close to $0.3c$.  
It is possible however that the AGN lacks a broad line region altogether. In such a case the constraint on the inclination angle is relaxed to allow angles smaller than $45\degr$ and the head advance speeds required will be lower, although still remaining relatively high.

The rather high advance speeds of the ends of the outer lobes suggest a relatively young age of 6~Myr for the Mpc-size outer double. Additionally, it may be noted that such high advance speeds have rarely been inferred for jets interacting with the ambient intergalactic medium: it is instead in the case of hotspots linked with restarting jets advancing in relic cocoons that such large advance speeds have been suggested \citep{saf08}.  

This model also expects that the approaching hotspot would have an increased flux density compared to the receding hotspot because of relativistic beaming of the hotspot emissions.  However, for inclination angles close to $45\degr$ along with advance speeds close to $0.3c$ the expected hotspot flux ratio is 4.3, which is well above the observed ratio of 1.6 (ratio of the brighter peak in the SE lobe to the NW hotspot peak). 

Here we point to the curious non-collinearity of the hotspots in the two outer lobes with the core as well as the multiple hotspots observed at the SE end. The SE hotspots are displaced to the east with respect to the axis formed by the core and the NW hotspot. This is in the same direction as that  suggested above for the axis flip of $30\degr$ from the direction of the inner offset lobes to the outer lobes.  We may understand the angular offsets between in the inner relic double and the outer hotspots, as well as the offsets between the hotspots at the two outer ends, by a model that invokes a clockwise rotation of the axis of the radio jets over time.   

From the angle of non-collinearity of about $10\degr$ between the NW and SE outer hotspots and the light travel time difference of nearly 2.5~Myr required to produce the seen asymmetry in extents of the two outer lobes we infer a clockwise rotational speed of about $4\degr$~Myr$^{-1}$ for the ejection axis of the radio jets. In this model that invokes episodic jets along with a rotating ejection axis, the inner double was created at an earlier epoch and is observed today as relic lobes defining that past axis orientation.  Subsequently the axis rotated and a new activity episode resulted in the observed hotspot at the NW end and a corresponding hotspot along that axis to the SE.   A southern extension from the current SE hotspot (Fig.~2) might be the remnant of that SE hotspot.  Continued episodic activity as the axis continued to rotate clockwise resulted in the bright hotspot at the SE end; the counterpart of this activity is yet to be observed at the NW end owing to light travel time delay in that receding lobe.  

This brightest hotspot in SE end is currently diminished in brightness because the jet in that lobe is already moving on to feed another hotspot at an advanced rotation angle; this sharing of the beam power may be a reason for this SE hotspot to be fainter compared to the expectations derived above for the inclination angle and advance speed.  Additionally, since the sound  crossing time for a 1~kpc hotspot is expected to be only $10^4$~yr, the hotspot flux will be variable on that timescale unless the beam is stationary in power and orientation.  For these reasons, it is difficult to relate the hotspot flux ratio to the Doppler
factor.

In the model proposed above the black hole spin axis rotates at a speed of  $4\degr$~Myr$^{-1}$ and at this speed it would have taken about 7~Myr to change from the direction of the inner relic double to the axis defined by the outer bright hotspots.  Adding to this the dynamical age of the giant radio source, we infer that the relic inner lobes are at least 13~Myr old.  This is consistent with the upper limit of $10^{8}$ yr derived above for the age of the inner double.  

The 14-195~keV X-ray luminosity reported for the source identified with the B0707$-$359 host galaxy in the Swift/BAT hard X-ray survey allows us to compare with the kinetic luminosity in the jets of GRG as a consistency check for the scenario considered above (that the active jets are presently oriented along the GRG and the source may be growing with a high advance speed of 0.3c).
We use the equipartition pressure derived for the inner lobes (we assume the same in the extended regions of the outer lobes) together with the dimensions of the outer lobes (625~kpc for the SE~lobe and 440~kpc for the NW lobe with a common width and depth of 100~kpc) and the estimated age for the GRG of 6~Myr to derive a jet kinetic luminosity of $1.9 \times 10^{44}$ erg s$^{-1}$ (for the SE jet) and $1.3 \times 10^{44}$ erg s$^{-1}$ (for the NW jet). These values compare well with the (14-195~keV) hard X-ray luminosity of $4.8 \times 10^{44}$ erg s$^{-1}$ for the AGN.

Some aspects of the scenario sketched above that we did not address are what causes the jets to rotate between the jets forming the relic lobes and the jets beginning to drive lobes in new directions and whether the jets could have been active throughout the rotation instead of discrete activity episodes that 
created the inner and outer lobe pairs. \citet{den02} discuss the causes of jet axis rotation that include on one hand change in accretion disk angular momentum 
(e.g. due to gas driven by minor merger) as the primary cause and on the other hand change in black hole axis (due to a past major merger) being the main driver. In either case the visible signs of the past events may not be strong and the estimated timescales for the jet axis rotation can range from half Myr to $10^{7}$ yr. As for the continued jet activity during the rotation, we cannot rule out the possibility and instead there may be some support for it in the
form of lack of clear trends as well as shallower than expected inner lobe spectral index distribution. Clearly, in the morphology of GRG B0707$-$359, we may be witnessing a consequence of jet rotation. Timescales, causes and effect on AGN activity all need to be further understood. B0707$-$359 may be a good candidate to explore these aspects of AGN physics.

In an altogether different scenario to explain the source structure there is the possibility that the inner and outer lobe pairs represent manifestations of two separate but close AGN whose axes are offset from each other (\citet{lal07}). As yet however there has been no 
convincing case of such a close pair of AGN both of which are also radio galaxies (3C75 known for its dual radio source structure has AGN cores 
that are separated by more than 7~kpc). There are several studies and programs however that are attempting to identify cases of close pairs of supermassive black holes and a few strong candidates have been reported where there are two VLBI cores \citep{fre12, rod06}. 

In the study of the extended structure in GRG B0707$-$359 presented here a few tests will be interesting and useful to carry out. For example if the inner double 
is indeed a relic source mapping it with higher resolution is not expected to reveal bright jets. Again, if the radio jet is aligned $45\degr$ to the line of sight then we expect in a high resolution optical/IR observation to see presence of broad emission lines in polarized flux. Also it will be useful to 
carry out VLBI observations of the core to check for presence of twin cores which are expected if the peculiar structure is because of two nearby AGN.

\section{Summary}

We have presented radio continuum observations in total intensity and polarization of the giant radio galaxy B0707$-$359.
The source has an unusual morphology in that two radio lobe pairs are recognized which are offset from each other at an angle of
30-degrees. Not only that there are two lobe pairs and that they are offset from each other, the new lobes are both of bottle-neck nature and protrude well out of the 250~kpc inner lobes to different distances. We have tried to understand how the source structure may have formed and we have presented 
arguments that lead to an internally consistent scenario for the formation of the source morphology as that of an AGN which has restarted its jets with the new and current jets propagating in a direction different from the original jets.  We infer that the AGN has restarted its jets in an offset direction; moreover, we also infer that  the new direction is not in the plane of the sky. The location of the giant radio galaxy axis defined by the hotspots, along an angle closer to the host minor axis rather than the major axis, supports a restarting and axis-change model for the formation of the structure we see in B0707$-$359. It is also consistent with a previous study suggesting a preference for giant radio galaxies to develop closer to host minor axes \citep{sar09}.

\section*{Acknowledgments}
ATCA is part of the Australia Telescope, which is funded by the Commonwealth of Australia for 
operation as a National Facility managed by CSIRO. We thank the anonymous referee for very helpful suggestions which have 
improved the paper.

\begin{figure*}
\includegraphics[width=120mm]{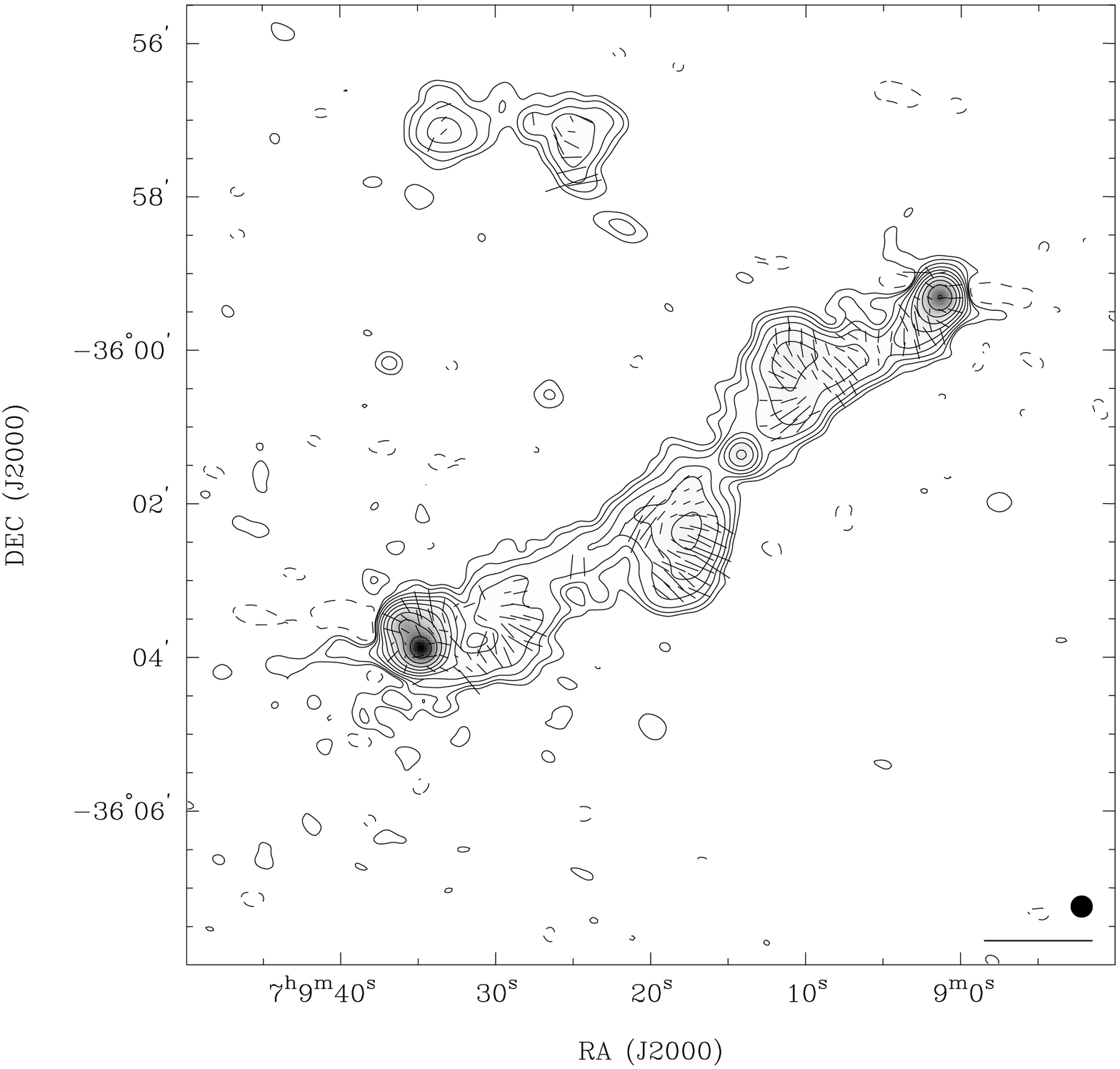}
\caption{ATCA 1384 MHz radio continuum - primary beam-corrected, total intensity image with a beam of FWHM 16\arcsec and total intensity contours at
-1, 1, 2, 4, 8, 16, 32, 64, 128, 256 $\times$ 0.5 mJy beam$^{-1}$. 
Vectors corrected for Faraday rotation have been overlaid. The vectors represent fractional polarisation (the scale bar represents 100 per cent polarisation). Rms noise in the map is 0.16~mJy/beam.}
\end{figure*}

\begin{figure*}
\includegraphics[width=120mm]{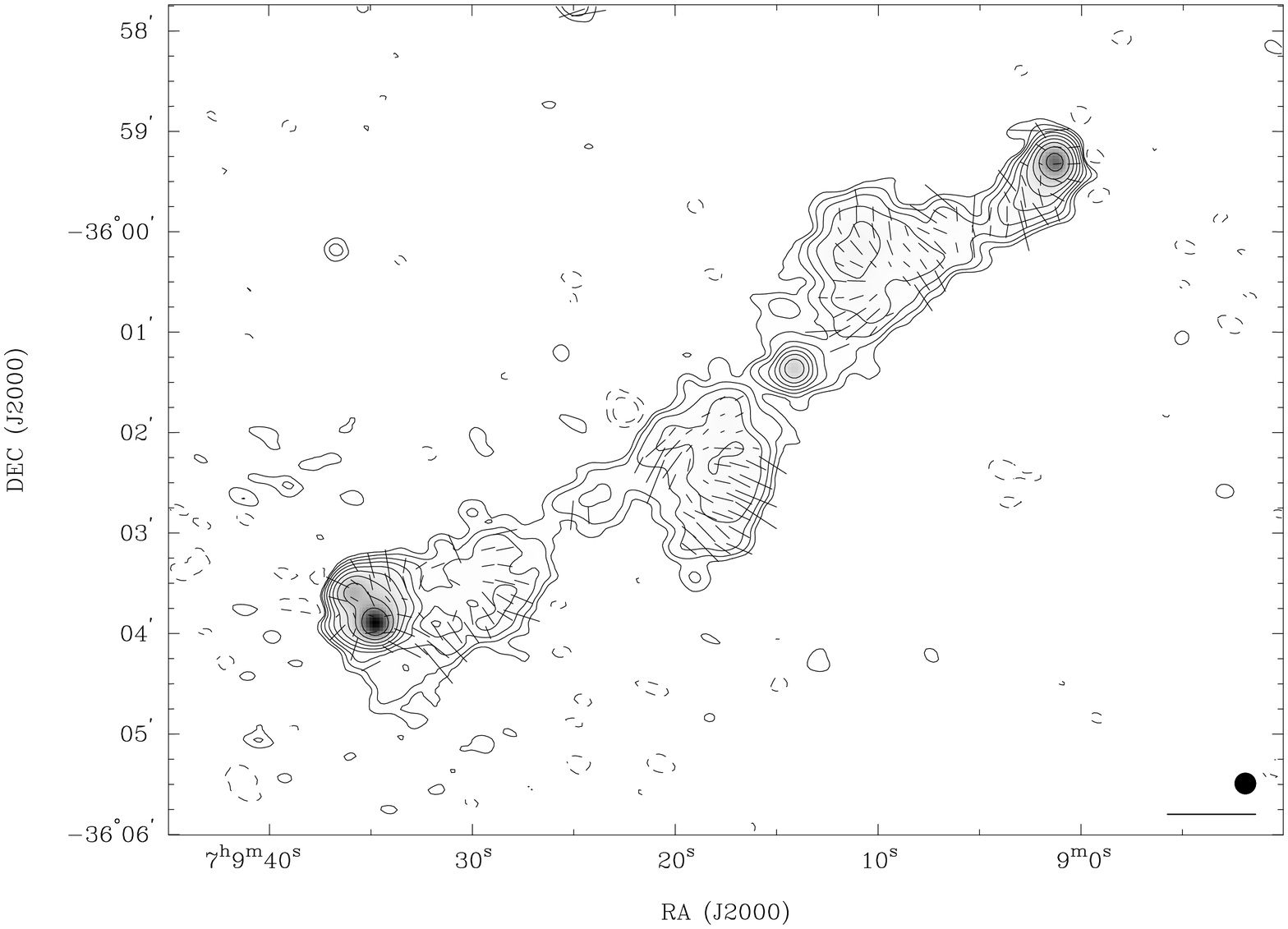}
\caption{ATCA 2368 MHz radio continuum - primary beam-corrected, total intensity image with a beam of FWHM 12\arcsec and total intensity contours at 
-2,- 1, 1, 2, 4, 8, 16, 32, 64, 128, and 256  $\times$ 0.4 mJy beam$^{-1}$. Vectors corrected for Faraday rotation have been overlaid. The vectors represent fractional polarisation (the scale bar represents 100 per cent polarisation). Rms noise in the map is 0.14~mJy/beam.}
\end{figure*}

\begin{figure*}
\includegraphics[width=120mm]{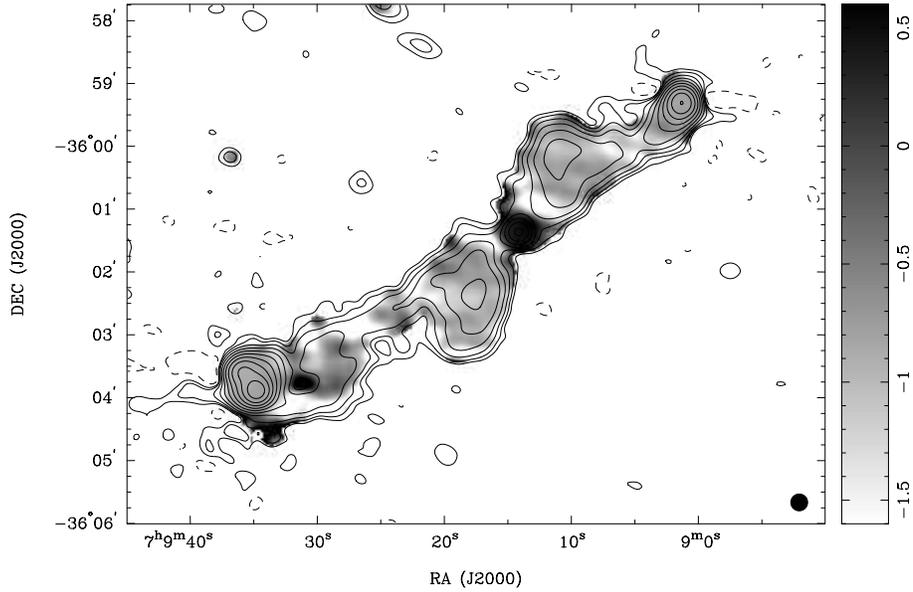} 
\caption{20cm - 13cm spectral index distribution shown in greyscale in the range -1.6 to +0.6 with a beam of FWHM 16\arcsec and overlaid with 1.4 GHz total intensity contours at -1, 1, 2, 4, 8, 16, 32, 64, 128, and 256 $\times$ 0.5 mJy beam$^{-1}$.}
\end{figure*}

 \begin{table}
 \caption{Journal of ATCA Observations at 1.4 and 2.3 GHz}
 \label{tbl:sourceprop}
 \begin{tabular}{lcc}
 \hline
 Array & Date              & Time \\
            &                       & (hr)    \\
 \hline
 6C$^*$     & 2000 Nov 11 &  9.09  \\
 1.5E$^*$  & 2000 Nov 22 &  10.63\\
 750C$^*$ & 2000 Dec 29 & 9.68  \\
 375$^*$    & 2001 Feb 27 & 10.27\\
 750D & 2004 Jun 14 & 3.79  \\
 6D      & 2004 Dec 07 & 9.98  \\
 \hline
 \end{tabular}

 \medskip
 Note -- $^*$ indicates observations with phase centre offset

\end{table}

\begin{table*}
 \caption{Source properties at 1.4 and 2.3 GHz}
 \label{tbl:sourceprop}
 \begin{tabular}{lcc}
  \hline
 Properties of the core                  & 1.4~GHz & 2.3~GHz  \\
 -Integrated flux density (mJy) & 44.6      & 53.0       \\
 -$\dagger$Spectral index   ($S_{\nu} \propto \nu^{\alpha}$)                       & \multicolumn{2}{c}{0.44} \\
  \hline
 Properties of the offset inner double& 1.4~GHz& 2.3~GHz  \\
 -Total flux [incl. core] (Jy)         & 0.49      & 0.33       \\
 -Power (W Hz$^{-1}$) [in the rest frame of the source] & $1.64\times10^{25}$ & $1.10\times10^{25}$ \\
 -$\dagger$Mean spectral index in N lobe & \multicolumn{2}{c}{-0.89} \\
 -$\dagger$Mean spectral index in S lobe & \multicolumn{2}{c}{-0.99} \\
 -Mean fractional polarisation for the N lobe (per cent) & 12.4 & 17.9 \\
 -Depolarisation ratio for the N lobe & \multicolumn{2}{c}{0.69} \\
 -Mean fractional polarisation for the S lobe (per cent) & 17.1 & 21.9 \\
 -Depolarisation ratio for the S lobe & \multicolumn{2}{c}{0.78} \\
 -Mean rotation measure in N lobe (rad m$^{-2}$) & \multicolumn{2}{c}{58.0} \\
 -Mean rotation measure in S lobe (rad m$^{-2}$) & \multicolumn{2}{c}{62.4} \\
 -Pressure of the N lobe (outer region) (Pa) & \multicolumn{2}{c}{2.03$\times10^{-14}$} \\
 -Pressure of the S lobe (outer region) (Pa) & \multicolumn{2}{c}{1.79$\times10^{-14}$} \\
  \hline
 Properties of the outer double   & 1.4~GHz & 2.3~GHz  \\
 -Total flux [incl. core] (Jy)        & 1.50      & 0.98       \\
 -Power (W Hz$^{-1}$) [in the rest frame of the source] & $5.05\times10^{25}$ & $3.31\times10^{25}$ \\
 -$\dagger$Spectral index at the N hotspot& \multicolumn{2}{c}{-0.80} \\
 -$\dagger$Spectral index at the S hotspot& \multicolumn{2}{c}{-0.83} \\
 -Lobe extent asymmetry &\multicolumn{2}{c}{1.47}\\
 -Mean rotation measure in N lobe (rad m$^{-2}$) & \multicolumn{2}{c}{60.3} \\
 -Mean rotation measure in S lobe (rad m$^{-2}$) & \multicolumn{2}{c}{63.3} \\
   \hline
 \end{tabular}

 \medskip
 Note: $\dagger$ indicates all spectral index values have been calculated using 16 \arcsec images
 at both 1.4~GHz and 2.3~GHz.
\end{table*}

\clearpage

\end{document}